\documentclass[11pt]{article}

\usepackage{wrapfig}
\usepackage{graphicx}
\usepackage{bbm}
\usepackage{color}

\usepackage{textcomp}
\usepackage[english]{babel}

\date{}

\begin{document}

\title{\Large\bf Entanglement and intra-molecular cooling in biological systems? -- A quantum thermodynamic perspective}

\author{\large Hans J. Briegel$^{1,2}$ and Sandu Popescu$^{3,4}$} \maketitle
\vspace*{-10mm} \begin{center}\small\emph{$^1$Institute for Theoretical Physics, University of Innsbruck, Technikerstra\ss e 25, Innsbruck, Austria\\
$^2$Institute for Quantum Optics and Quantum Information of the Austrian Academy of Sciences,
Technikerstra\ss e 21a, A-6020 Innsbruck, Austria\\
$^3$H.H. Wills Physics Laboratory, University of Bristol, Tyndall Avenue, Bristol BS8 1TL, U.K.\\
$^4$Hewlett-Packard Laboratories, Stoke Gifford, Bristol BS12 6QZ, U.K.}
\end{center}

\begin{abstract}
We discuss the possibility of existence of entanglement in biological systems. Our arguments centre on the fact that biological systems are thermodynamic open driven systems far from equilibrium. In such systems error correction can occur which may maintain entanglement despite high levels of de-coherence.  We also discuss the possibility of cooling (classical or quantum) at molecular level.
\end{abstract}

\section{Introduction}

The notion of entanglement plays a central role in quantum physics. It is a key concept in quantum information theory, with applications in quantum computation and communication \cite{Ben00,Ste98}, and considerable effort has been devoted to identifying physical systems in which entanglement can be created and exploited for information processing purposes \cite{Zol05}. From a broader perspective, entanglement and the concepts of quantum physics play a fundamental role for our understanding of Nature: The occurrence of entanglement in any system indicates that we are on a terrain of Nature where classical concepts are likely to be quite insufficient for its proper understanding.

Experience shows that entanglement - other than the one that is there because it is a property of the ground state of a system and protected by brute force via an energy gap - is extremely fragile and easily destroyed by noise. Laboratories work very hard to produce such entanglement, e.g. by isolating single ions in traps and by cooling them close to their motional ground state or, more generally, by complete control and manipulation of matter on the atomic level. Hence most researchers believe that entanglement will be hard to find in natural (i.e. uncontrolled) systems outside the laboratories, not to mention biological systems, which operate at room temperature and which are complex, noisy and ``wet" (for a recent review see e.g.\ \cite{Wis07}). Here however, we would like to make the case that contrary to the standard view, entanglement might actually exist and play a role also in biological systems; one should be aware of this and look for signatures of entanglement in such systems.

In fact, possible evidence for quantum coherence (though not entanglement) in photosynthetic systems has been recently 
reported in \cite{Eng07,Lee07} and studied theoretically in a number of papers \cite{Moh08,Ple08,Ola08,Reb09}.

We want to emphasize from the outset that this is not a research paper in the sense that we do not have any concrete results to prove that persistent, controlled entanglement exists in biological systems. We will however try and present circumstantial evidence for it.

Our main point is that the intuitions about the fragility of entanglement that are used for dismissing the possibility of entanglement in biological systems are misleading because they generally ignore a fundamental fact, namely that biological systems are open driven systems. This opens many possibilities that, as far as we know, have not yet been carefully considered.

Since the issue is so important, we feel that it would be a great mistake to dismiss the possibility of biological entanglement without a much more careful investigation. The scope of our paper is to call for such a vigorous program of research.

Related to the problem of biological entanglement is possibility of intra-molecular cooling, that is, the possibility that parts of a molecule are actively cooled relative to the environmental temperature.  This process is in fact much more general than the generation of entanglement. In fact intra-molecular cooling might occur also in instances when there is no quantum coherence whatsoever. As such, the probability that intra-molecular cooling actually takes place in biological systems is far larger than the probability of existence of controlled quantum coherent phenomena and entanglement that form the main subject of the present paper.

\vspace*{-0.15cm}
\section{Live or dead entanglement?}

To start with, we should discuss more carefully what kind of entanglement are we looking for.

That entanglement, and, more generally, coherent quantum effects always exist - at some level - in all systems (including biological ones) is quite clear.  After all the laws of quantum physics enter on the level of quantum chemistry, determining the structure and energy spectra of the molecules and their interactions. Coming to biology, most scientists would probably share the view that quantum mechanical effects play a role, but only an indirect one: Quantum mechanics would thus be responsible for the molecular basis or substrate, while the biological functionality of the molecules can be explained by classical statistical physics, combined with the principles of molecular Darwinism \cite{Hug99}.

This makes it clear that, even before asking whether entanglement exists or not, we need to better define what we are actually talking about, that is: what kind of entanglement?  In the following, we would like to distinguish three different kinds of entanglement that we expect to play a role in biology; the same classification holds also for any quantum coherent processes that may occur in biological systems.

One has to say from the beginning that the boundaries between theses different types of entanglement are fuzzy. The classification of some phenomena is clear-cut, while for others one may argue whether it is of type 1 or type 2, etc.; nevertheless, we believe this classification to be essential when proceeding to study the possibility of biological quantum effects.
\bigskip

The three types of entanglement are:
\begin{itemize}
\item entanglement of basic constituents
\item dead entanglement
\item live entanglement.
\end{itemize}

\bigskip
\noindent
{\bf Entanglement of basic constituents.}
Think of any atom or molecule; these elementary systems contain a lot of entanglement. Indeed, all their electrons are entangled, the protons and the neutrons in the nucleus are entangled, and so on. The existence of this sort of entanglement is obvious and in most cases trivial.

As always the boundary cases may be more interesting. For example, while the existence of a delocalized electron state in a benzene molecule is also quite trivial, finding delocalized states that extend over much larger molecules may have important consequences.

A similar situation occurs in condensed matter systems, in which the ground state is typically highly entangled, too. The role of this entanglement e.g. in quantum magnets or super-conducting materials has attracted a lot of attention recently \cite{Ami08}. The phenomenon of entanglement is thus not restricted to the microscopic domain of very few particles, but it occurs also in macroscopic bodies and even at finite temperatures.

The entanglement in these systems (which could be called ``static entanglement'') has many interesting facets. In any case, this is not the type of entanglement we are concerned with in the present paper.

\bigskip
\noindent
{\bf Dead entanglement.}
Dead entanglement occurs in molecules that have biological origin or occur in biological cells. However the occurrence of this kind of entanglement does not require metabolic processes to function. As such, these molecules can be taken out of the cell, and they will continue to work.

Here we are talking about systems that are generally in thermal equilibrium. When some appropriate external perturbation comes, it takes them out of equilibrium, some coherent phenomenon takes place and it quickly dies out. A paradigmatic example of this, outside the biological context, would be a piece of an optical fiber. When a photon comes it propagates through the optical fiber and gets out at the other end. However, during the rest of the time, the piece of optical fiber just stays there, say on an optical table, and nothing happens to it. A similar example, taken from biology, would be the more complicated molecule of chlorophyll, which absorbs energy (in form of a photon) that then propagates (in form of an exciton) from one centre/part of the molecule to another part, until it reaches the reaction center.  Again, in the absence of the photon, the molecule is at equilibrium and nothing interesting happens to it.

There are some key words and properties that we generally expect to be associated with this type of entanglement: \emph{Incidental}, \emph{Side effect}, \emph{Short time},  \emph{May have biological functionality}, and \emph{May be evolutionary selected}.

{\it Incidental:} The main characteristic of the process may not require entanglement, or coherence, but in a particular implementation of the process they may just occur. For example, when a photon propagates through an optical fiber, maintaining polarisation coherence is not necessary. It so happens however that present day optical fibers are so good that polarisation coherence is maintained. But as far as functionality is concerned, which is transmitting the light, this is not important.

{\it Side effect:} Entanglement may be just some sort of side effect of the process. That is, it may always accompany a given process, but not play any role.

{\it Short time:} Generally these phenomena are short time because this external perturbation produces some modification but then the environment immediately brings it down to equilibrium. These are phenomena that may typically take pico- or femto-seconds. As a matter of fact, if you work on a very short time scale, there is always some quantum coherence, because it requires some time to die out, to de-cohere.

On the other hand, one must also be aware that although the absolute time scales involved in these phenomena are very short, this doesn't necessarily mean that entanglement/coherence does not play a significant role. Indeed, a relevant time scale is that of the duration of the process itself. If entanglement/coherence is present during the whole process, then it may plays a significant role; otherwise its role is most probably irrelevant.

{\it Biological functionality and evolution:} In some instances, the entanglement {\it may} have some biological functionality, and it {\it may} have been that this type of quantum coherence was evolutionary selected. It is also possible that the although the entanglement is just a side effect without biological functionality, the primary effect that leads to it was biologically selected. That is, other, important, things were selected and evolved, and with them the entanglement, but just as an accompanying effect.  As an example, we can give here photosynthesis again.

The key word here is ``may''. That is, in the case of this type of entanglement/quantum coherence, while evolution may occur, it is not a sine-qua-non condition for its very existence.

By no means are ``dead'' entanglement/quantum coherence non-interesting phenomena. On the opposite. They are extremely interesting and very complicated: Although conceptually they are the same, there is an enormous difference between say, propagation of a photon through an optical fiber and propagation of an exciton through chlorophyll. Showing that even such ``simple'' quantum effects actually take place in biological systems is a great challenge. In fact at present all the experimental work on biological quantum effects is focussed exactly on this type of processes.

\bigskip
\noindent
{\bf Live entanglement.}
The defining property of this type of biological entanglement is that it exists only while metabolic processes take place. In other words, it exists only as long as the system is actively maintained far from thermal equilibrium, i.e. in {\bf open, driven systems far from equilibrium}. When the metabolism stops and the system reaches equilibrium, this type of entanglement disappears.

The key properties that we expect this class of phenomena to posses are:
\emph{It is persistent}, \emph{It is dynamically controllable}, \emph{It has biological functionality}, \emph{It is evolutionary selected}.

{\it Persistency:} By their very nature, these processes are such that as long as they are active, entanglement/coherence is maintained. This is in fact the purpose of the entire process. From this point of view they are fundamentally different from dead-entanglement phenomena, in which the entanglement appears only as a transitory phenomenon.

{\it Dynamical control:} In the case of dead entanglement, such as, if confirmed, propagation of an exciton through a chlorophyll molecule, the details of the process are governed by the structure of the molecule itself. This is a structural and therefore relatively static parameter. On the other hand, in processes that are driven far from equilibrium, changes in the way in which the process is driven can immediately alter the characteristics of the process, and hence of the entanglement.

{\it Biological functionality and evolutionary selection:}
These are, of course, much more complicated processes than the ones that give rise to dead entanglement. Unless the live entanglement has biological functionality, evolution most probably could not have arrived at it.

Again, all the above are properties that we expect this class to posses. This is not to say, of course, that there could be no exceptions. For example it is not impossible that some process leads to persistent entanglement as a side effect with no biological functionality. All we say is that this seems to us highly improbable.

To conclude this section, we want to emphasize once more that we do not have any (experimental) evidence, at present, that this type of entanglement (or similarly: quantum coherence) exists. The very scope of this paper is to investigate the possibility that such entanglement actually exists.

We would also like to mention that we are by no means suggesting the possibility of entanglement at very large scale - such as super-positions of brain states leading possibly to quantum computation in the brain, etc.. This seems to us virtually impossible and here we fully agree with the sceptical view expressed in Ref. \cite{Wis07} (see also \cite{Koc06}).
What we are interested in is persistent and controllable entanglement with presumably biological function, at the level of bio-chemical processes.

\section{Why should we look for entanglement in biological systems?}

At first sight, biological systems, which are warm and wet, seem very unlikely places to look for entanglement. It is our claim however that if entanglement (of the type we described above) is to be found anywhere in Nature outside physics labs, it is inside living beings.

To stabilize entanglement is generally difficult and requires complex setups - obtaining and maintaining entanglement in a laboratory is not an easy task. Of course, some instances of naturally occurring entanglement could exist in nature, such as a piece of nonlinear crystal that in sunlight may produce entangled photons by down-conversion. But such occurrences are purely accidental and probably very rare. On the other hand, biology itself could be a driving force: if entanglement turns out to be useful, biological evolution could select for this. In fact it is quite probable that entanglement offers advantages. The point is that the number of possible entangled states is so much larger than that of non-entangled ones that it is, in fact, inconceivable not to find entangled states that will offer advantages. Indeed, the possibilities offered by entanglement are much richer than those offered by non-entangled mixtures: If the efficiency of a certain biological process depends on the state that occurs in the system, the state space over which evolution can optimize that process is much larger if one allows entangled states. This is essentially a probability argument. To conclude, from this perspective, if entanglement is to be found anywhere in Nature, then in biological systems. Incidentally, the only place where we have non-trivial entanglement, i.e. in some laboratories, it is of ultimate biological origin!  It is only at the end of a long evolution that the required complexity for producing controlled entanglement has been achieved.

On the other hand, there is a caveat: Evolution takes a long time and it is quite possible that even though entanglement is useful, nature has not succeeded yet in exploiting it.

\section{Open driven systems and entanglement}

Living organisms depend on permanent consumption of energy in the form of food (or photons, as in plants). Different from e.g. some solid-state material, a living cell cannot be described as an isolated system. It continuously exchanges particles with its environment, and with them energy and entropy. In the language of thermodynamics, biological systems are open driven (quantum) systems, whose steady state is far away from thermodynamic equilibrium.  As we mentioned in the introduction, this fact has major implications for the issue of the presence of entanglement.

To begin with, the fact that in open driven systems entanglement can exist at room temperature is absolutely clear, once one realizes that every quantum physics laboratory is such a system. So, having established that, as a matter of principle, controllable entanglement can exist in room temperature systems, the question is only one of scale and complexity. Do we need sophisticated lasers and large fridges (which work at room temperature but cool a subpart of them) or can they be present in the small scale of a living cell? After all, there are numerous studies that show that analogues of large scale, man-made machines (ratchets, rotors, etc.) do actually exist on a bio-molecular level (for a review see e.g.~\cite{Bro06}). In what follows, we will give a number of specific examples that seem suitable to be scaled down.

Before going to specific examples however, it is important to understand why open driven systems make a difference. The reason is that such systems can perform error correction. Decoherence introduces noise into systems and increases their entropy. On the other hand, open driven systems have, by definition, access to a source of free energy and can use it to get rid of the errors. In fact for the issue we consider, namely merely producing and stabilizing a particular entangled state, one doesn't even need full quantum error correction, i.e. an error correction protocol that can stabilize any arbitrary superposition of states (in a given Hilbert subspace) \cite{Ste96,Cal96}. The error correction we require here is, computationally speaking, trivial - the stabilization of a single state. Nevertheless, entropically the task is similar, i.e. the entropy continuously produced by noise has to be removed from the system.

Coming back to biology, we note that probably the most striking characteristic of biological systems is that they are error correcting systems - a dead animal starts decomposing in a matter of hours so it must be that while living there are continuous error correction processes going on which maintain the body. So once we realize that error correction takes place in any living organism, whether or not it is enough to stabilize entangled states becomes a matter of scale not one of principle.

\subsection{Toy models I}

As a first example, consider the famous procedure used to produce entanglement in present day optical labs, namely parametric down conversion. A typical parametric down-conversion experiment uses a laser to produce a stream of photons that are directed towards a nonlinear crystal. Upon impinging onto the crystal, some of these photons generate pairs of polarization-entangled photons.

The entanglement in the experiment described above is totally dependent on the fact that we are dealing with an open driven system: The laser is powered by a power source; when the source is turned off the entanglement disappears.

\begin{figure}[ht]
\begin{center}
\includegraphics[width=11cm]{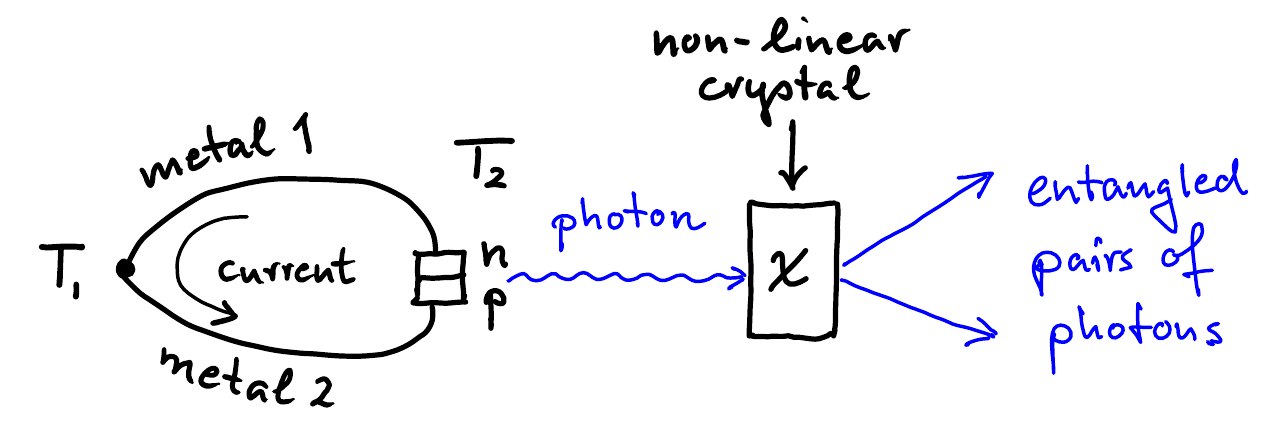}
\caption[]{\label{SimpleDevice} Simple entanglement-generating device (see text).}
\end{center}
\end{figure}

The parametric down-conversion experiment discussed above requires complex equipment including a laser and a power source. Is such complicated equipment necessary? The answer is no - there is no fundamental principle of nature that requires complex equipment. The following simple device (see Fig.~\ref{SimpleDevice}) could do the same thing.

Suppose we have a thermo-electric element - two simple wires, each made of a different metal - that is connected to a simple light-emitting-diode LED. The LED itself is a very simple device as well - two semiconducting crystals, each with a different impurity, joined together (a so called ``n-p junction"). If one of the joints of the thermo-electrical element is heated relative to the other, e.g. by a simple flame, the LED will produce a light-beam. When the output of the LED is directed to a non-linear crystal entangled photons are produced by parametric down-conversion.

Note that there is something quite remarkable about this simple device. (a) It is quite robust and has no moving parts. (b) The entire system is entirely driven by a temperature difference, thus, the source of free energy is not coherent, and neither is the transmission of current in the wires a coherent process (see also \cite{Ple02}). Furthermore, (c), the junction and the crystal are both at room temperature. This simple device is able to create something highly quantum mechanical, such as entanglement! How can that be? Of course, it has to do with the (coherent) interaction that goes on inside the crystal, when a photon is absorbed and converted into a pair of photons with lower energy. The main point to be observed is that this process is embedded into a hot and rather noisy environment, and it is only due to the continuous pumping of the system that the entanglement generating interaction in the crystal can be exploited.

We may further imagine that the photons in the entangled pair are further directed to two atoms that absorb them. Upon absorbing the photons, the atoms become entangled. If the atoms are subjected to noise, the entanglement could be subsequently destroyed; whether any entanglement can survive for a longer time is a question of how quick is the decoherence versus the rate of pumping new entanglement into the atoms by subsequent entangled photon pairs.

\begin{figure}[ht]
\begin{center}
\includegraphics[width=9cm]{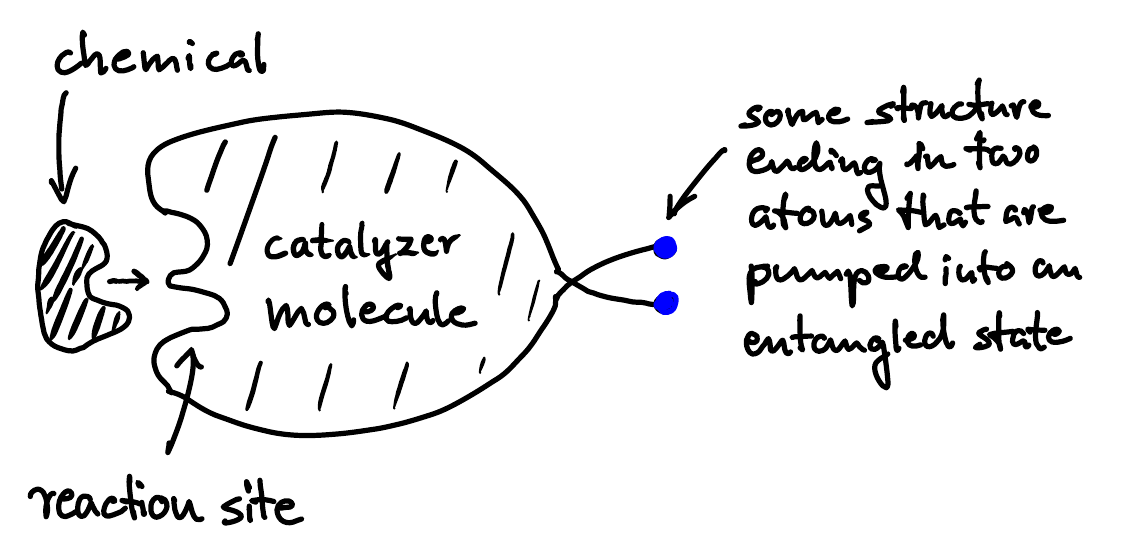}
\caption[]{\label{MolecularDevice} Hypothetical molecular device that is capable of creating entanglement by a molecular pumping process.}
\end{center}
\end{figure}

A molecular analogue (see Fig.~\ref{MolecularDevice}) would be a chemical process where a large molecule (such as a protein) serves as a catalyzer for an exothermic chemical reaction. We assume that the chemical reaction takes place at a certain functional centre (``docking site") of the catalyzer molecule. The chemical process may lead to a transfer of free energy along the molecule; the free energy transfer may be conveyed by various channels, for example by excitons, phonons, or electric current (electron displacement). This energy transport over the molecule need not be coherent and the molecule may be at room temperature. What matters is that, at some other site of the catalyzer molecule, the energy may be converted by some nonlinear process, similar to the one that occurs in the non-linear crystal of the parametric down-conversion, into two entangled modes that pump entanglement into, say, some receiver atoms.

The important point is here that the two atoms, depicted in blue in Fig.~\ref{MolecularDevice}, will not be entangled if we stop driving the system, i.e. if we stop the supply of the reactant chemical. This emphasizes the difference between passive (static) entanglement, as between electrons in an atom, and the dynamic equilibrium entanglement.

\subsection{Toy models II. Conformational changes and time-dependent Hamiltonians.}\label{ConformChanges}

In the previous section we emphasized that set-ups that produce entanglement need not be very complex, As a matter of fact however biological processes are actually very complex and sophisticated. A very important process that occurs in many instances is that of controlled conformational changes in proteins \cite{Alb08}. These processes could lead to controlled time-dependent Hamiltonians, with obvious implications for possible coherent quantum processes and existence of entanglement.

A conceivable process is the following.

\begin{figure}[ht]
\begin{center}
\includegraphics[width=11cm]{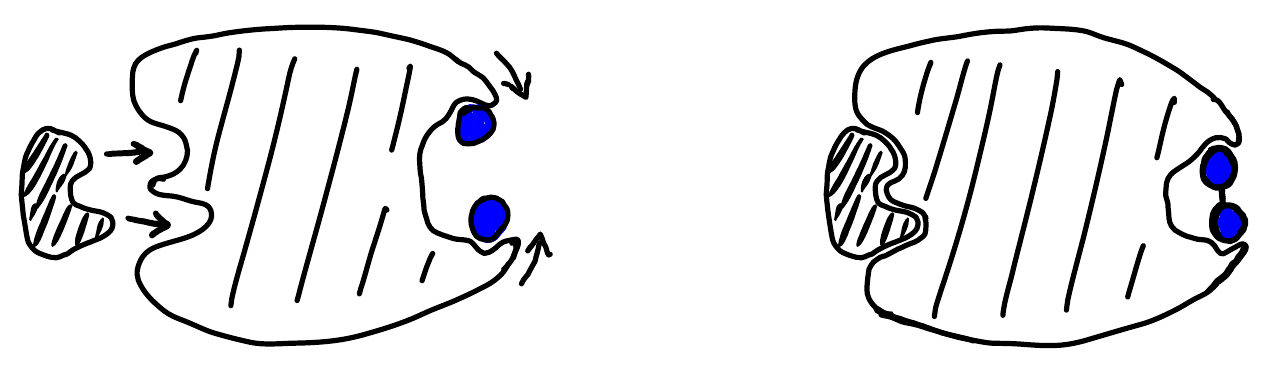}
\caption[]{\label{AllostericDevice} Entanglement of two atoms (blue) in a molecule, induced by a stream of reactant chemicals which dock to the catalyzing molecule, leading to a conformation change (see text).}
\end{center}
\end{figure}

\noindent A chemical process that occurs at one binding site of a protein can lead to a configuration change of another site of the same protein \cite{footnote2}. This conformation change may then lead to an interaction e.g. between two atoms that would otherwise be in a shielded site of the molecule (see Fig.~\ref{AllostericDevice}). The two atoms depicted in blue in Fig.~\ref{AllostericDevice} are not interacting in the ``rest" state of the protein; they are at thermal equilibrium and non-entangled. When the conformation change occurs they are brought together and start interacting. The interaction takes the atoms out of their equilibrium state and may entangle them. The entanglement survives until de-coherence kills it and the atoms reach a new thermal equilibrium state. Then protein reverts to the rest conformation and the atoms revert to the rest mode and then the process starts again. The point here is that although both thermal equilibrium states may be non-entangled, the state of the atoms during the transition time (which may be relatively long) may well be entangled.  Again, to complete the cycle the protein needs to be supplied with free energy; if we stop driving the system, i.e. if we stop the supply of reactant chemicals then the atoms simply reach thermal equilibrium and remain there and entanglement is lost. As in the previous section, this emphasizes the difference between passive (static) entanglement, as between electrons in an atom, and dynamic entanglement that exists due to the (non-equilibrium) process of conformational changes.

In the above example we described a very simple cycle. It is conceivable however that more complicated cycles could exist, with many different conformational changes occurring sequentially. Such a process may then implement a more intricate time-dependent Hamiltonian corresponding to what in quantum information is called a ``sequence of gates".

\subsection{Toy models III. Intra-molecular cooling.}

Thinking of different ways of obtaining controllable entanglement in the laboratory, with an eye to possible biological implementation has led us to the idea of intra-molecular cooling. This process is far more general than the generation of entanglement and needs not be associated with entanglement at all. In fact intra-molecular cooling might occur also in instances when there is no quantum coherence whatsoever. As such, the probability that intra-molecular cooling actually takes place in biological systems is far larger than the probability of existence of controlled quantum coherent phenomena and entanglement that form the main subject of the present paper.

Cooling, if it could actually be achieved at molecular level, would have obvious benefits. One example is when there are many possible reaction channels and, when the site is cooled, a preferred channel is selected. Another example is increased efficiency of catalysis: Many proteins act as catalyzers with very high specificity. They have active sites in the shape of cavities, which bind only molecules have that fit precisely into the cavity like a hand in a glove. At high temperature the protein vibrates stronger and the shape of the cavity is deformed which leads to a decline in the efficiency of catalysis. Obviously, if the active site could be cooled this would be a benefit.

Regarding the potential role of cooling in biological systems, it should be realized that many organisms have indeed the ability of cooling parts of their body.  For example our body temperature is kept largely stable at a temperature around 37 \textcelsius, even when we live in hot climate. This is only possible since our body has a built-in cooling system. The only question is therefore not whether or not cooling exists in biological organisms but at which scale. Does it exist only at large scale - at the scale of the whole organism, or at scale of internal organs - or is it present all the way down to molecular level? Given the potential advantages for intra-molecular cooling, it is not inconceivable that such mechanisms were evolutionary selected. In fact, it is even possible that cooling at molecular level is much stronger than at the scale of the whole body. The reason for this is two fold. First in smaller systems, the coupling of parts of the system with the environment may be much weaker than of other parts so some parts, the ``inside of the fridge" could be rather well isolated from the environment. Second, there may be a lot of free energy available (from chemical reactions), which will be channeled preferentially within the molecule.

Set-ups that allow cooling can be surprisingly simple. An explicit example is the so called ``algorithmic cooling" \cite{Bau05, Bra05, Sch99} which could be realized even on a molecular level. In algorithmic cooling, we have in mind that some atoms at a given site, presumably an active but protected site, might be cooled.  A very simple example of algorithmic cooling \cite{Bra05} involves only three qubits (i.e. two-level systems). In their original example Brassard \emph{et al.} \cite{Bra05} considered the qubits to be nuclear magnetic moments, but any other physical system (such as vibrational levels of atoms) could be used. In this example one qubit can be cooled while the two other qubits get warmer - very much like the back spiral of an ordinary refrigerator. From these two warmer qubits heat is dissipated into the environment and the process continues, achieving persistent cooling of the first qubit.

In the algorithmic cooling process, the interactions between the qubits are time dependent, driven by an external mechanism.  Essentially, this can be realized by a (short) sequence of pair wise couplings between the qubits. In practice, this could be realized by, say, changing the relative position of atoms according to a predetermined sequence, something that could be realized by a sequence of conformational changes in a protein. Consider for example a protein molecule with two active sites, similar to the one in Fig.~\ref{AllostericDevice}. The active site on the left is where we supply free-energy (by some catalytic reaction) which is required for driving the cooling process. The active site on the right (where the blue atoms in Fig.~\ref{AllostericDevice} are) is the ``main active site" that we want to cool. The chemical reaction at the active site on the left drives the cooling process at the active site on the right.

Of course, it is impossible to measure directly the temperature of an active site of a protein. However one possible signature for the presence of a cooling process would be the following: As mentioned above the typical rate of reaction for biological catalytic processes is the following. At low temperature the reaction is slow because the reactants move slower. Then the rate increases with the temperature. However, after a certain temperature the reaction rate starts decreasing because the reaction site becomes deformed due to vibrations and the reactants don't fit in it anymore.

\begin{figure}[ht]
\begin{center}
\includegraphics[width=8cm]{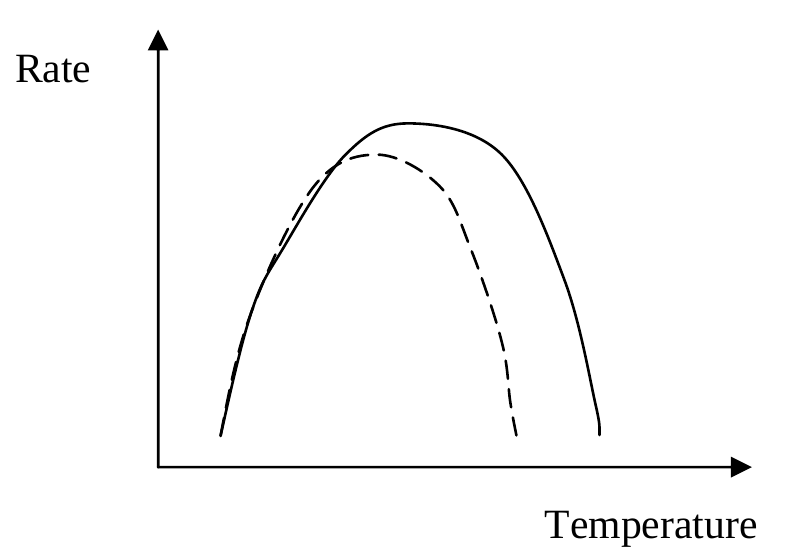}
\caption[]{\label{AllostericEnzymeControl} Potential control of enzyme activity through intra-molecular cooling (see text).}
\end{center}
\end{figure}

Consider now the activity of the site subjected to cooling, described in Fig.~\ref{AllostericEnzymeControl}. The temperature axis of the graph represents the temperature of the environment of the molecule (the temperature of the cell, or of the liquid that surrounds the cell.  The two curves describe the rate of this reaction in two cases - when the protein is supplied with free-energy and when not. Recall that we supply free-energy in the form of chemical reactants that react on the active site at left. When there are no reactants to drive the cooling process, the main active site is at the same temperature as the environment. On the other hand, when there is a supply of reactants the main active site is cooled relative to the environment and the rate of the reaction it catalyzes remains high at higher environmental temperatures. Note however that there should be no difference between the two cases at low temperatures because then the rate is simply limited by the slow movement of the molecules of interest, i.e. they enter less frequently in the main active site.

Of course, there could be many reasons why the rate of reaction at the main active site increases when a reaction occurs at the other site of the molecule - the molecule's configuration could simply be optimized for the case when both reactions occur simultaneously. However, the very specific temperature dependence described above could be a good indicator.

\subsection{Toy models IV. Reset mechanisms.}

\begin{figure}[ht]
\begin{center}
\includegraphics[width=11cm]{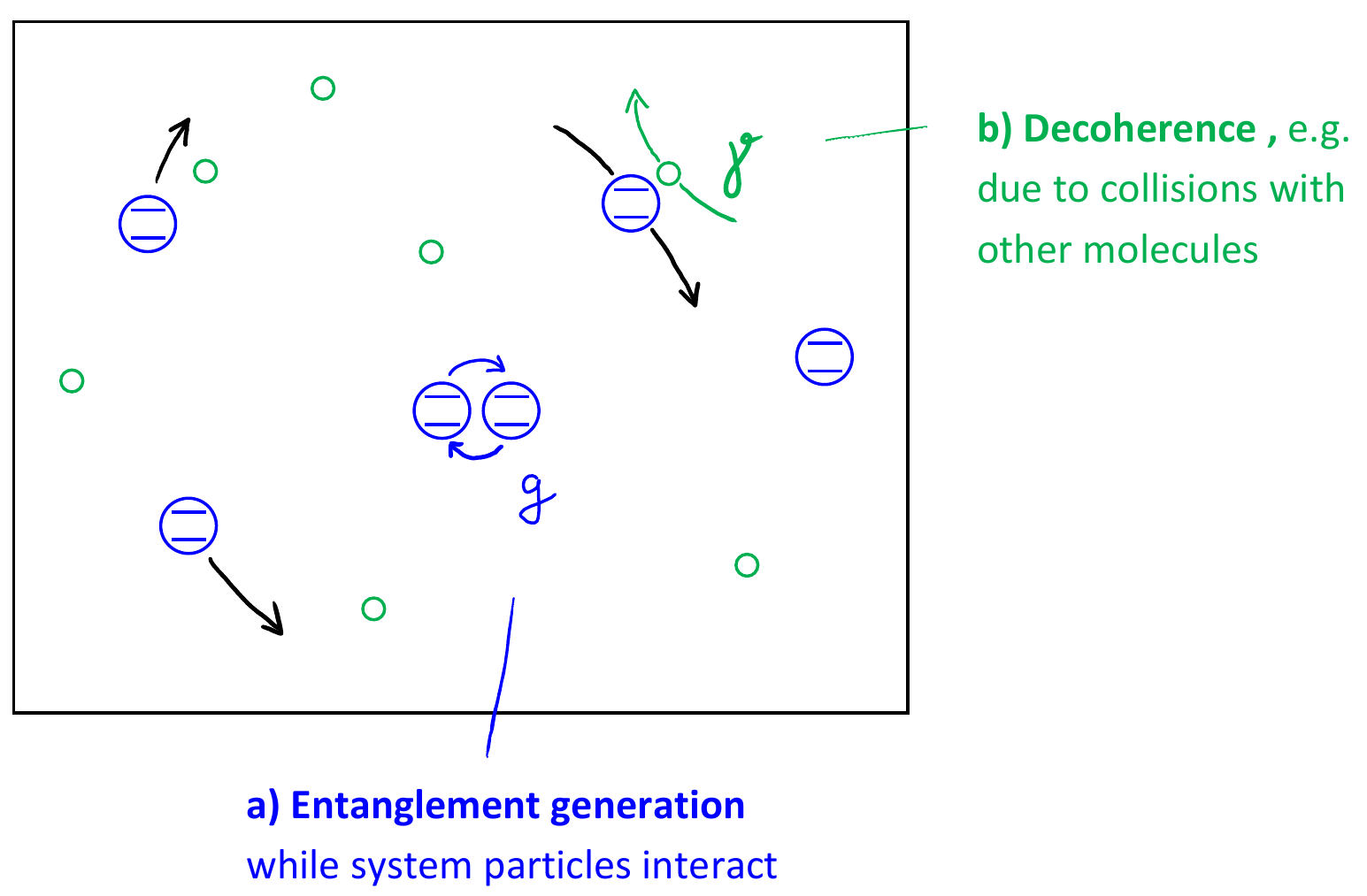}
\caption[]{\label{SpinGas} a) Illustration of a spin gas (blue): Particles move on classical trajectories, but each particle carries an internal quantum mechanical degree of freedom, such as a spin. Upon collision, the spin degrees of freedom get entangled.  b) Collisions of the gas particles with other particles (buffer gas, green) lead to de-coherence.}
\end{center}
\end{figure}

Our last example is not derived from a macroscopic device, but from the study of another instance of an open driven quantum system away from thermal equilibrium.  This is a gas-type system - specifically a spin gas - that was studied in \cite{Cal05}. This particular system is relatively simple and it is accessible to quantitative analysis.
The reason why it is instructive to study this example has to do with the role of de-coherence in such a system, which couples to the individual gas particles and thus quickly destroys any transient entanglement. It is thus clear that no static entanglement can occur in such a system.
Let us consider a simple gas cell, where gas particles move on classical trajectories, but have some internal structure which is described by quantum mechanics (see Fig.~\ref{SpinGas}). The interaction between these particles is capable of entangling their internal degree of freedom. As a concrete realization, we may e.g. imagine ultra-cold atoms with two internal hyperfine states $|0\rangle$ and $|1\rangle$, or molecules carrying a nuclear spin with two possible states. For the sake of the argument, suppose that we have full knowledge about the entire interaction history of the gas particles. As was found in a series of studies, the quantum states generated under such a simple dynamics, can already display a number of interesting and highly non-trivial entanglement properties. These include states with a high persistency of entanglement and states that are universal resources for quantum computation. This illustrates that even a system as simple as this (toy-model of a) gas can exhibit highly non-trivial and complex quantum effects.

Consider now that, as in any real scenario, these states will be exposed to de-coherence, which will quickly destroy the entanglement. A possible mechanism for de-coherence could be collisions of the gas particles with the environment consisting e.g. of other species of particles - a ``buffer gas" - within the gas cell (see Fig.~\ref{SpinGas}). The effect of such additional interactions, even if they are weak, will be devastating and in general no entanglement will survive in steady state.

\begin{figure}[ht]
\begin{center}
\includegraphics[width=11cm]{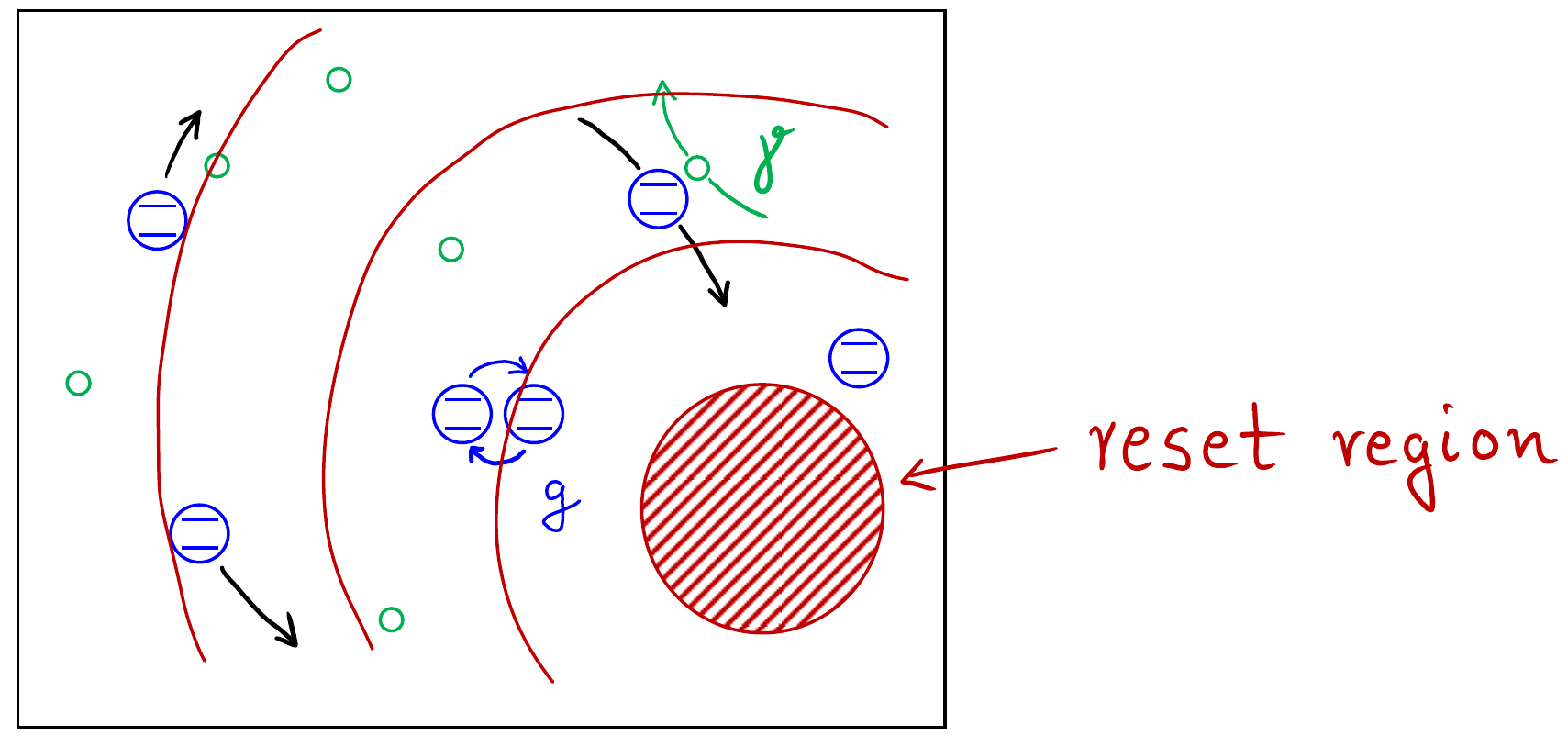}
\caption[]{\label{Reset} Particles that pass through the reset region indicated by a red spot will be reset in some standard internal state of low entropy. Particles that cross the spot will thus leave the spot in a pure state $|\chi\rangle = \alpha |0\rangle + \beta |1\rangle$. As a consequence, entanglement will build up around the spot and persist -- in dynamic equilibrium -- despite of de-coherence.}
\end{center}
\end{figure}

However, as was shown in \cite{Har06}, there is a simple mechanism that can in principle sustain the entanglement in the gas in the presence of de-coherence, without introducing entanglement by itself! Imagine that, when particles enter a certain region in the cell, indicated by a red spot in Fig.~\ref{Reset}, they are reset in some pure state $|\chi\rangle = \alpha |0\rangle + \beta |1\rangle$. This could be realized by various mechanisms, for example by an interaction with some local structure in the cell.

Such a reset mechanism will have two effects. It destroys existing entanglement between this particle (i.e. its spin) and the other particles in the gas, but it also destroys this particle's
entanglement with the environment! On first sight, this does not appear to be very constructive, but it has the effect that the particles that leave the light spot are in a pure state, and, if two of such particles collide later on (e.g. in the vicinity of the spot), they are capable of creating fresh entanglement! A detailed analysis shows indeed that, for a specific choice of parameters this (toy-model of a) gas can have a steady state, where entanglement persists in the cell. Depending on the mean free path of the particles, the regions of persisting entanglement may be confined to the vicinity of the reset region or they may extend over the entire gas cell.

While the details of this process are not of interest here, it should be mentioned that there are two regimes, one of vanishing and one of non-vanishing entanglement, and there is a sharp transition between them. Entanglement can be sustained if both the coherent interaction strength and the reset rate are sufficiently large. Furthermore, the particles need not be reset to a pure state; it suffices if they leave the region in a mixed state of sufficiently low entropy/high purity.

The main purpose of the preceding discussion of the spin gas was to demonstrate that the possibility of entanglement is not confined to highly controlled systems at very low temperatures: Simple reset processes allow entanglement to persist also in a hot and noisy environment! Even though this is a simple model, it is conceivable that similar processes could play a role also in biological systems, where various analogues of reset processes exist.

\section{Conclusions}

Biological systems are of extraordinary complexity and diversity. As such, at the moment we don't know where to start searching for entanglement and/or molecular cooling, be it an experimental search or a theoretical one. Furthermore, the specific toy models presented here are almost surely with very little direct relevance in biology. However, our goal here is far more limited - it is to argue that the presence of controlled entanglement with biological functionality cannot be discounted automatically, without a careful study. Indeed, although our specific toy models may well have very little direct relevance, we are confident that the processes we described (entanglement pumping, resetting, etc.) are to be found in a way or another; the same applies to the idea of molecular cooling. Ultimately, the power of biological evolution coupled with the fact that biological organisms are open, driven systems, may open the door for many unexpected quantum phenomena. Similarly, they also open the door to highly non-trivial thermodynamic phenomena.

\bigskip
{\em Note added:} Since the first version of this paper was put on ArXiv, a number of preprints have appeared which investigate the role of quantum entanglement in specific biological scenarios, including e.g. photosynthesis \cite{Wha0905} and the chemical compass mechanism for magnetoreception \cite{Cai09,Rie09}.
The possibility of dynamic entanglement generated by conformational changes of molecules, as described in Sec. \ref{ConformChanges}, has in the meantime been
studied quantitatively in \cite{CaiPopescuBriegel08}.


\end{document}